\documentclass{ifacconf}

\usepackage{graphicx}      
\usepackage{natbib} 
\usepackage{amsmath}
\usepackage{amssymb}     
\usepackage{subfigure}   
\usepackage[latin1]{inputenc}
\usepackage[T1]{fontenc}
\newtheorem{remark}{Remark}  
       
\begin{document}
\begin{frontmatter}

\title{Solar energy production: \\ Short-term forecasting \\ and risk management 
} 


\author[First,Fourth,Fifth]{C\'{e}dric Join} 
\author[Second,Fourth]{Michel Fliess} 
\author[Third,Sixth]{Cyril Voyant}
\author[First]{Fr\'{e}d\'{e}ric Chaxel}

\address[First]{CRAN (CNRS, UMR 7039), Universit\'{e} de Lorraine, BP 239, \\ 54506 Vand{\oe}uvre-l\`{e}s-Nancy, France \\ (e-mail: \{cedric.join, frederic.chaxel\}@univ-lorraine.fr)}
\address[Second]{LIX (CNRS, UMR 7161), \'Ecole polytechnique, 91128 Palaiseau, France (e-mail$:$ Michel.Fliess@polytechnique.edu)}
\address[Third]{SPE (CNRS, UMR 6134), Universit\`{a} di Corsica Pasquale Paoli, 20250 Corte, France (e-mail: voyant@univ-corse.fr)}
\address[Fourth]{AL.I.E.N. (ALg\`{e}bre pour Identification \& Estimation Num\'{e}riques), 24-30 rue Lionnois, BP 60120, 54003 Nancy, France \\ (e-mail: \{michel.fliess, cedric.join\}@alien-sas.com)}
\address[Fifth]{Projet Non-A, INRIA Lille -- Nord-Europe, France}
\address[Sixth]{H\^{o}pital de Castelluccio, Unit\'{e} de Radioth\'{e}rapie, BP 85, \\ 20177 Ajaccio, France}

\begin{abstract}             
Electricity production via solar energy is tackled via short-term forecasts and risk management. Our main tool is a new setting on time series. It allows the definition of ``confidence bands'' where the Gaussian 
assumption, which is not satisfied by our concrete data, may be abandoned. Those bands are quite convenient and easily implementable. Numerous computer simulations are presented.
\end{abstract}


\end{frontmatter}

\section{Introduction}
\subsection{Generalities}
The following lines by \cite{reikard} provide an excellent introduction to our subject: \textit{The increasing use of solar power as a source of electricity has led to increased interest in forecasting radiation over short time horizons. Short-term forecasts are needed for operational planning, switching sources, programming backup, and short-term power purchases, as well as for planning for reserve usage, and peak load matching.}  There are many approaches as summarized by \cite{traperoetal}: \textit{The diversity of solar radiation forecasting methodologies can be classified according to the input data and the objective forecasting horizon. For instance, NWP (Numerical Weather Prediction) models, which are based on physical laws of motion and conservation of energy that govern the atmospheric air flow, are operationally used to forecast the evolution of the atmosphere from about 6 h onward. Although NWP models are powerful tools to forecast solar radiation at places where ground data are not available, many near-surface physical processes occur within a single grid box and are too complex to be represented and solved by equations. Thus, NWP models cannot successfully resolve local processes smaller than the model resolution. Satellite-derived solar radiation images are a useful tool for quantifying solar irradiation at ground surface for large areas, but they need to set an accurate radiance value under clear sky conditions and under dense cloudiness from every pixel and every image \dots These limitations have placed time series analysis as the dominant methodology for short-term forecasting horizons from 5 min up to 6 h. } See \cite{kleissl} for a slightly different 
standpoint.

Diverse viewpoints on time series have of course been employed. See, \textit{e.g.}, \cite{bacher,diagne,duchon,lauret,martin,reikard,traperoetal,v1,v2,energy,yang}, and the references therein. We follow here another \emph{model-free} setting\footnote{See \cite{ijc} for the importance of the model-free viewpoint in control. It might worthwhile in our context to stress that this approach has also been successful for the renewable energy production (\cite{jama,edf}).}  (\cite{perp,esaim,ipag,douai,agadir}). With respect  to solar energy production they have already been compared to techniques stemming especially from persistence and from artificial neural nets by \cite{paris} and by \cite{esp}. Let us emphasize that our techniques are quite far from today's dominant viewpoint on time series (see, \textit{e.g.}, \cite{meuriot} and the references therein).

\subsection{Forecasting and risk}
According to a theorem due to 
\cite{cartier} the following additive decomposition holds for any time series $X$ under quite weak assumptions:
\begin{equation}\label{decomposition}
\boxed{X(t) = E(X)(t) + X_{\tiny{\rm fluctuation}}(t)}
\end{equation}
where
\begin{itemize}
\item the \emph{mean}, or \emph{average}, or \emph{trend}, $E(X)(t)$ is quite smooth,
\item $X_{\tiny{\rm fluctuation}}(t)$ is quickly fluctuating.
\end{itemize}
The decomposition \eqref{decomposition} is unique up to a ``small'' additive quantity. Our short-term forecast techniques are based on a local mathematical analysis of $E(X)(t)$ (\cite{perp,agadir}), which is inspired by recent advances in the field 
of estimation. They yield good results and are quite easy to implement. Their application to solar energy by \cite{paris}  and by \cite{esp} do not necessitate contrarily to most other approaches  \emph{big data}, \textit{i.e.}, large historical data.  Any type of forecast is always approximate. Here, according to Formula \eqref{decomposition}, the quick fluctuations $X_{\tiny{\rm fluctuation}}(t)$ explain to a large extent this discrepancy.  This inherent \emph{risk} does not seem to have been seriously investigated in the literature on solar energy although it plays obviously a key r\^{o}le in the energy production. We follow (\cite{troyes}) and exploit the viewpoint on \emph{volatility} developed by \cite{douai,agadir}. Classic normality tests show that our concrete time series are not related to Gaussian processes. We replace therefore the well known \emph{confidence intervals}, which do not make much sense in this situation, by the \emph{confidence bands} which bear some similarity with the famous \emph{Bollinger bands} in \emph{technical analysis} (\cite{bollinger}).\footnote{Compare with \cite{trapero}.}This might be an important advance in risk analysis.
\begin{remark}
See, \textit{e.g.}, \cite{willink} for a most rewarding account on confidence intervals.
\end{remark}

\subsection{Organization of the paper}
The paper by \cite{troyes} in this conference already presents a summary of our approach to time series. Therefore this material will not be repeated here. It gives more room to Section \ref{exper} where 
\begin{itemize}
\item a type of volatility is considered,
\item confidence bands are defined,
\item we report quite numerous numerical experiments which are based on real meteorological data. 
\end{itemize}
Some thoughts about solar energy forecasting and risk management are discussed in Section \ref{conclusion}.

\section{Numerical experiments}\label{exper}
\subsection{Presentation}
Write $X(t)$ the solar irradiance at time $t$. Those data are given by measurements every minute in Nancy, France, during the year 2013. In order to simplify the presentation of our computer calculations, we only utilize here the months of February and June.

\begin{figure*}
\begin{center}
\subfigure[Monthly view]{
\resizebox*{6.05cm}{!}{\includegraphics{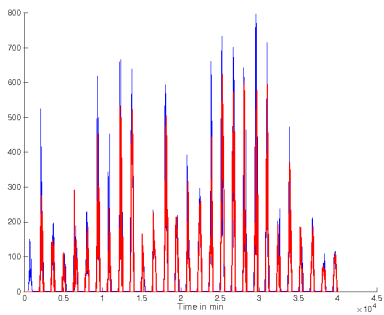}}}%
\subfigure[Zoom of \ref{figfp}-(a)]{
\resizebox*{6.05cm}{!}{\includegraphics{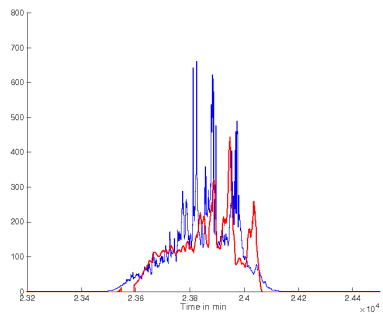}}}%
\subfigure[Zoom of \ref{figfp}-(a)]{
\resizebox*{6.05cm}{!}{\includegraphics{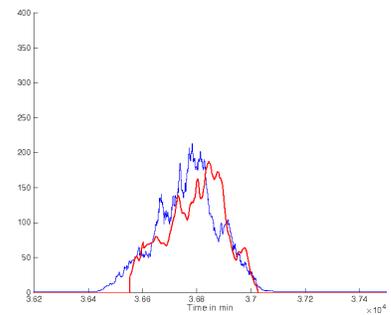}}}%
\caption{February : irradiance (blue) and its prediction (red)}%
\label{figfp}
\end{center}
\end{figure*}

\begin{figure*}
\begin{center}
\subfigure[Monthly view]{
\resizebox*{6.05cm}{!}{\includegraphics{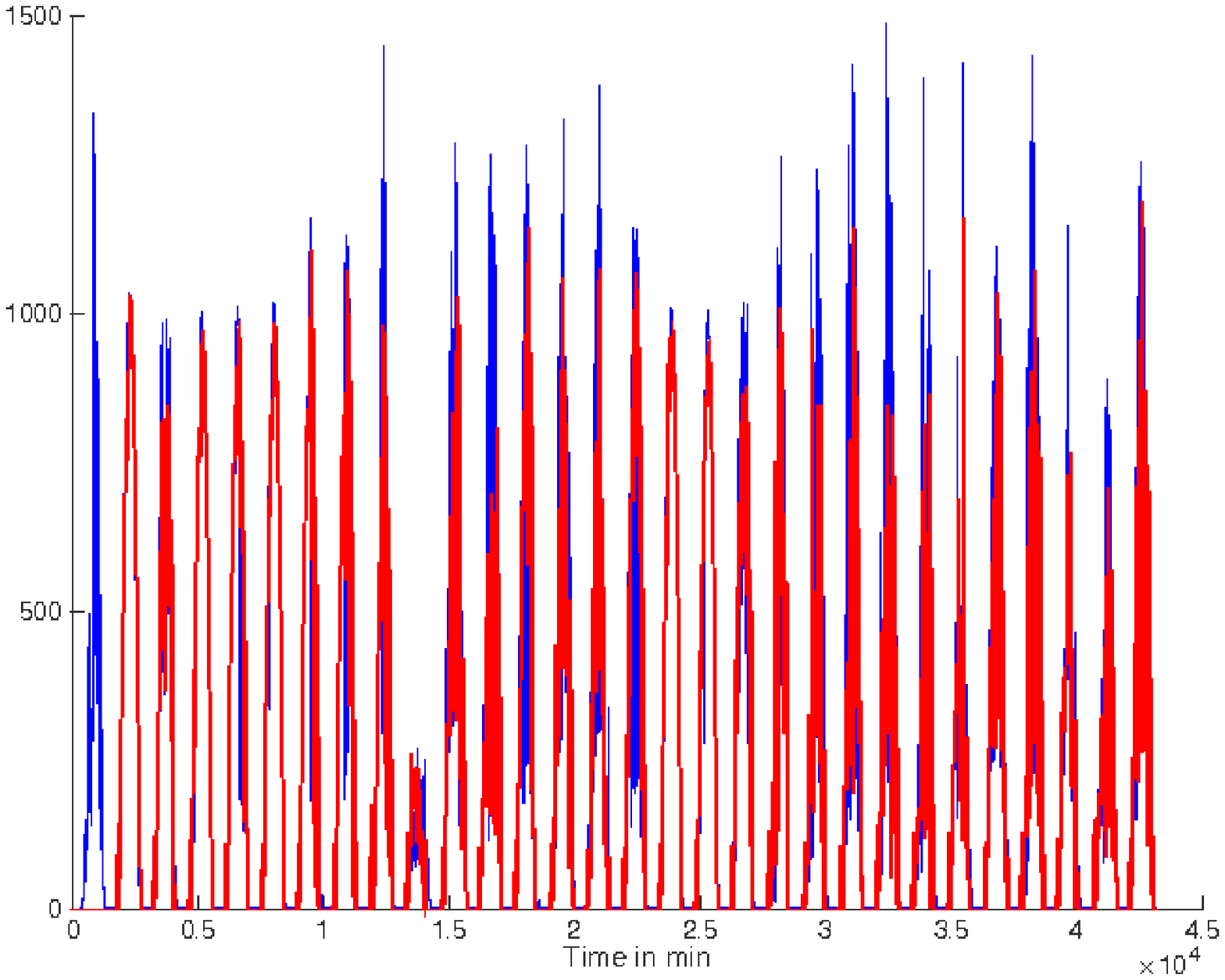}}}%
\subfigure[Zoom of \ref{figjp}-(a)]{
\resizebox*{6.05cm}{!}{\includegraphics{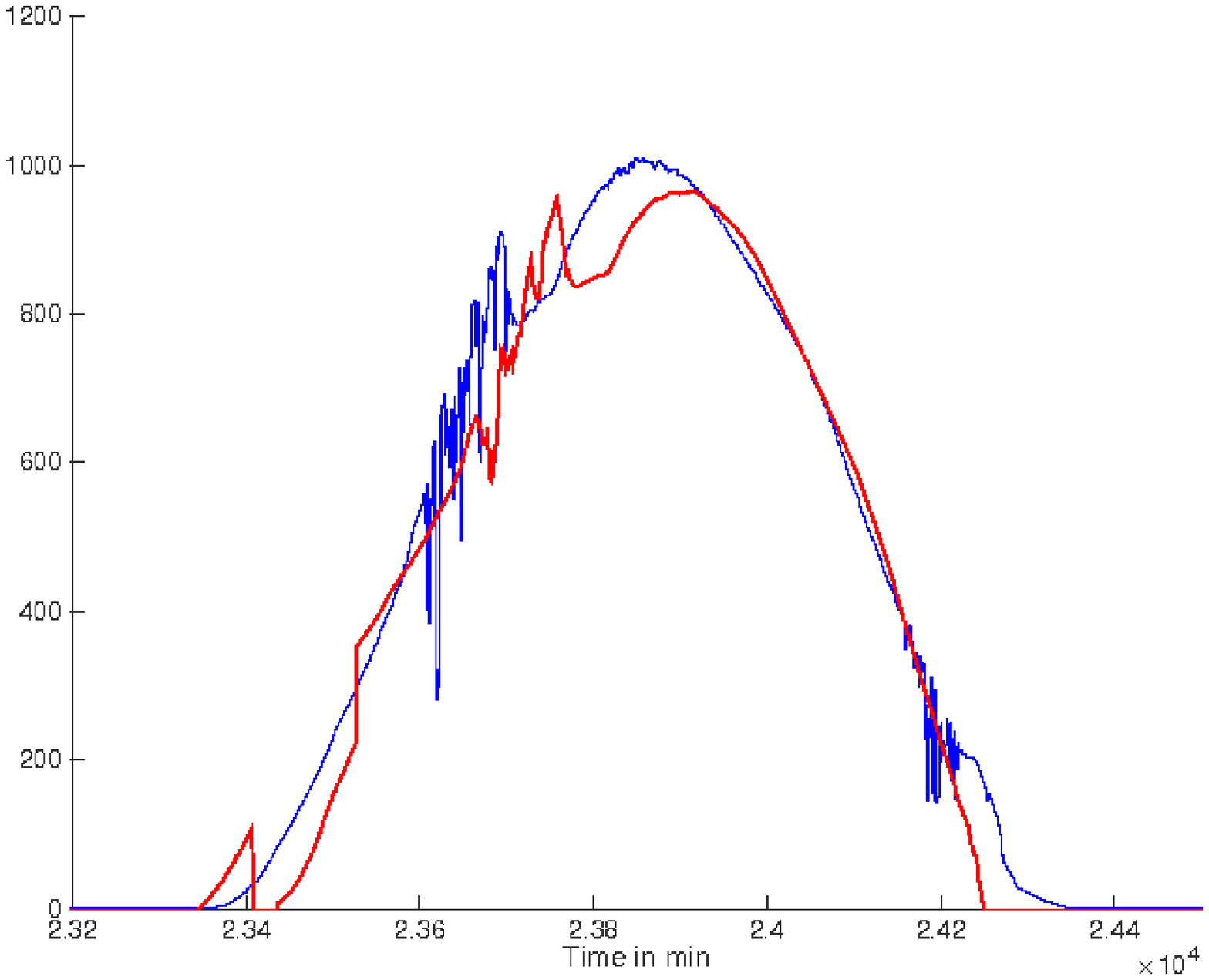}}}%
\subfigure[Zoom of \ref{figjp}-(a)]{
\resizebox*{6.05cm}{!}{\includegraphics{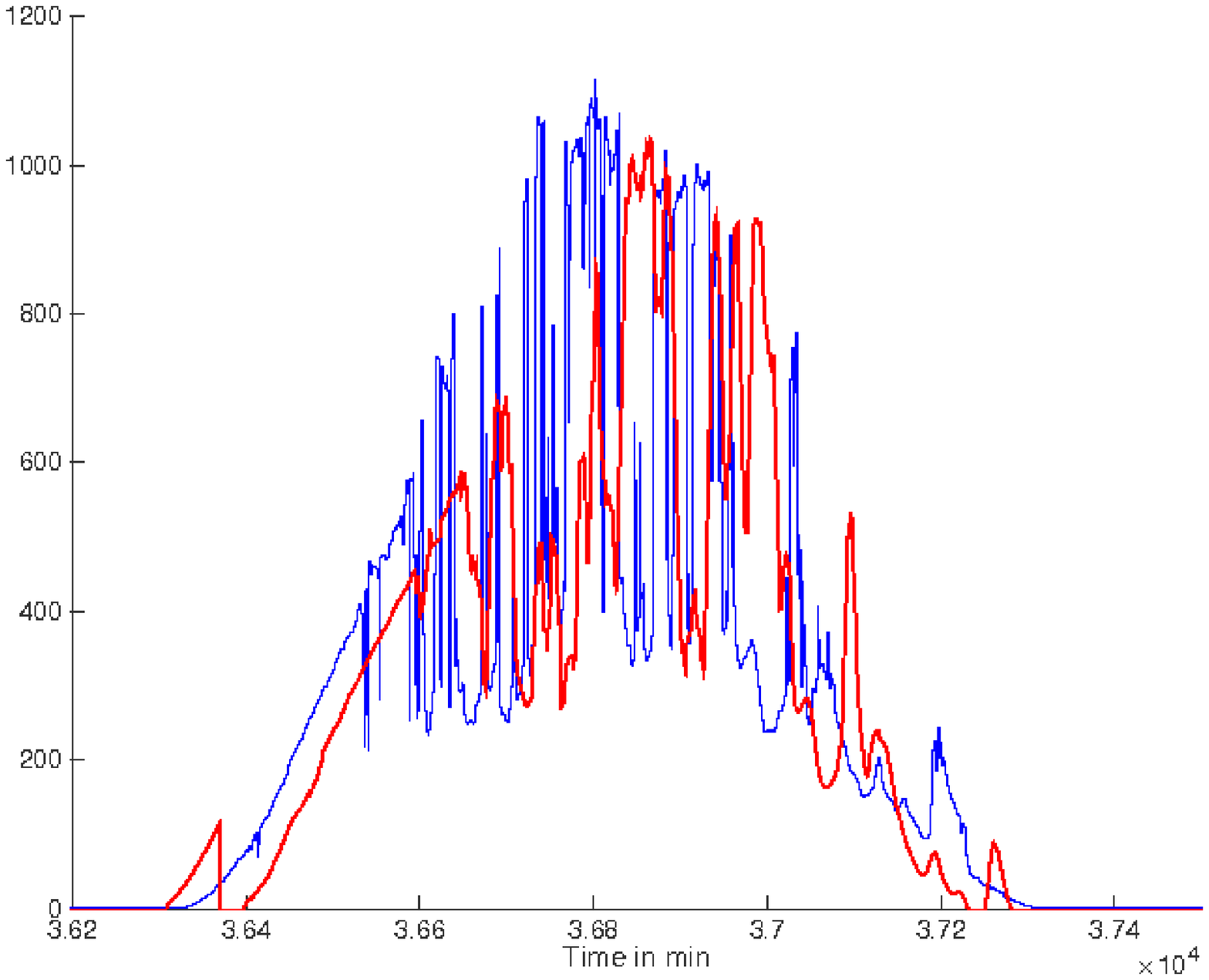}}}%
\caption{June : irradiance (blue) and its prediction (red)}%
\label{figjp}
\end{center}
\end{figure*}

Write $X_{\text{pred$60$}}(t )$ the forecast 1 h ahead of $X(t)$ as obtained by \cite{paris}.\footnote{Here again, we do not reproduce the calculations.} The results  in Figures \ref{figfp} and \ref{figjp} are borrowed from the same reference.

\subsection{Volatility}
\begin{figure*}
\begin{center}
\subfigure[Monthly view]{
\resizebox*{6.05cm}{!}{\includegraphics{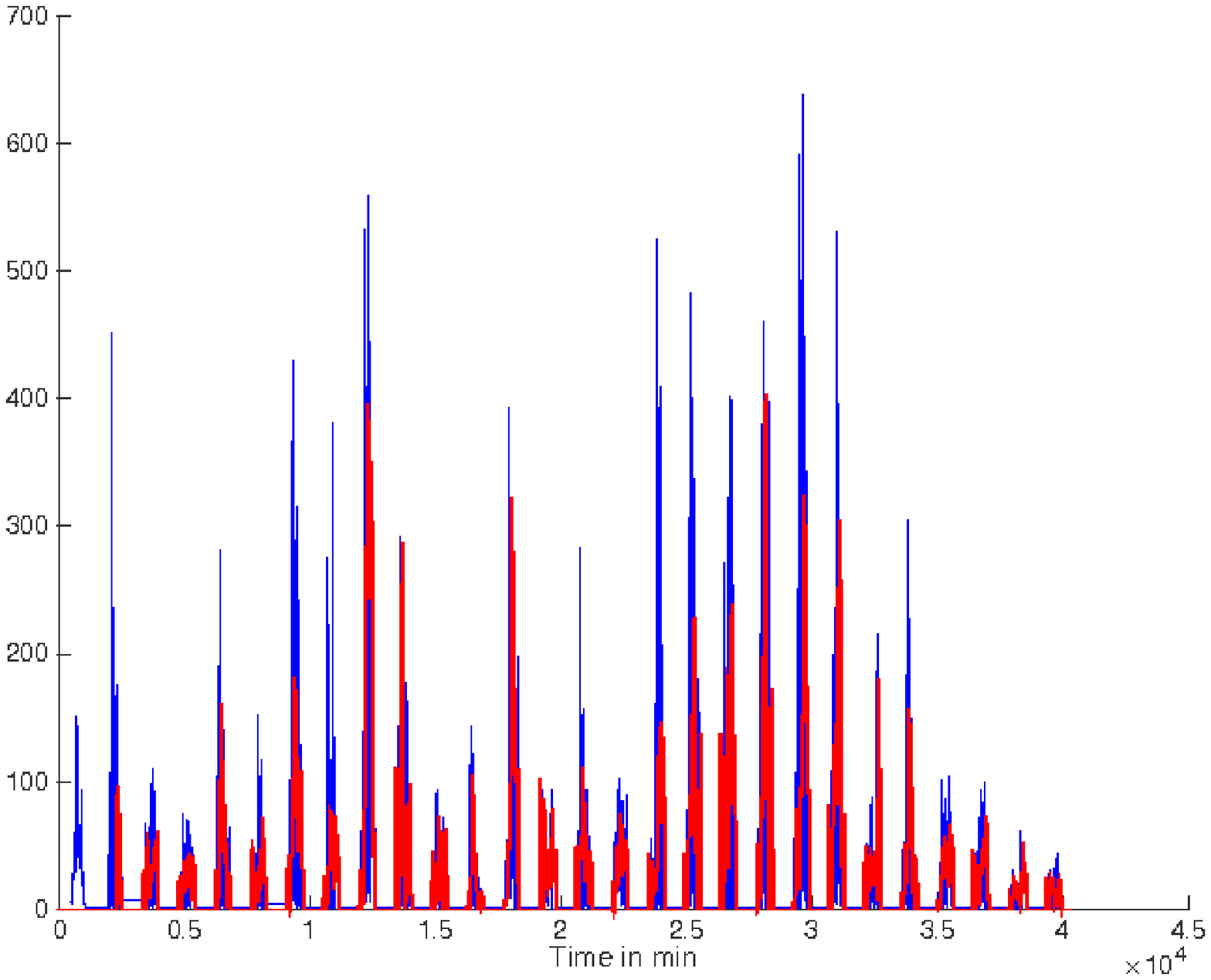}}}%
\subfigure[Zoom of \ref{figfv}-(a)]{
\resizebox*{6.05cm}{!}{\includegraphics{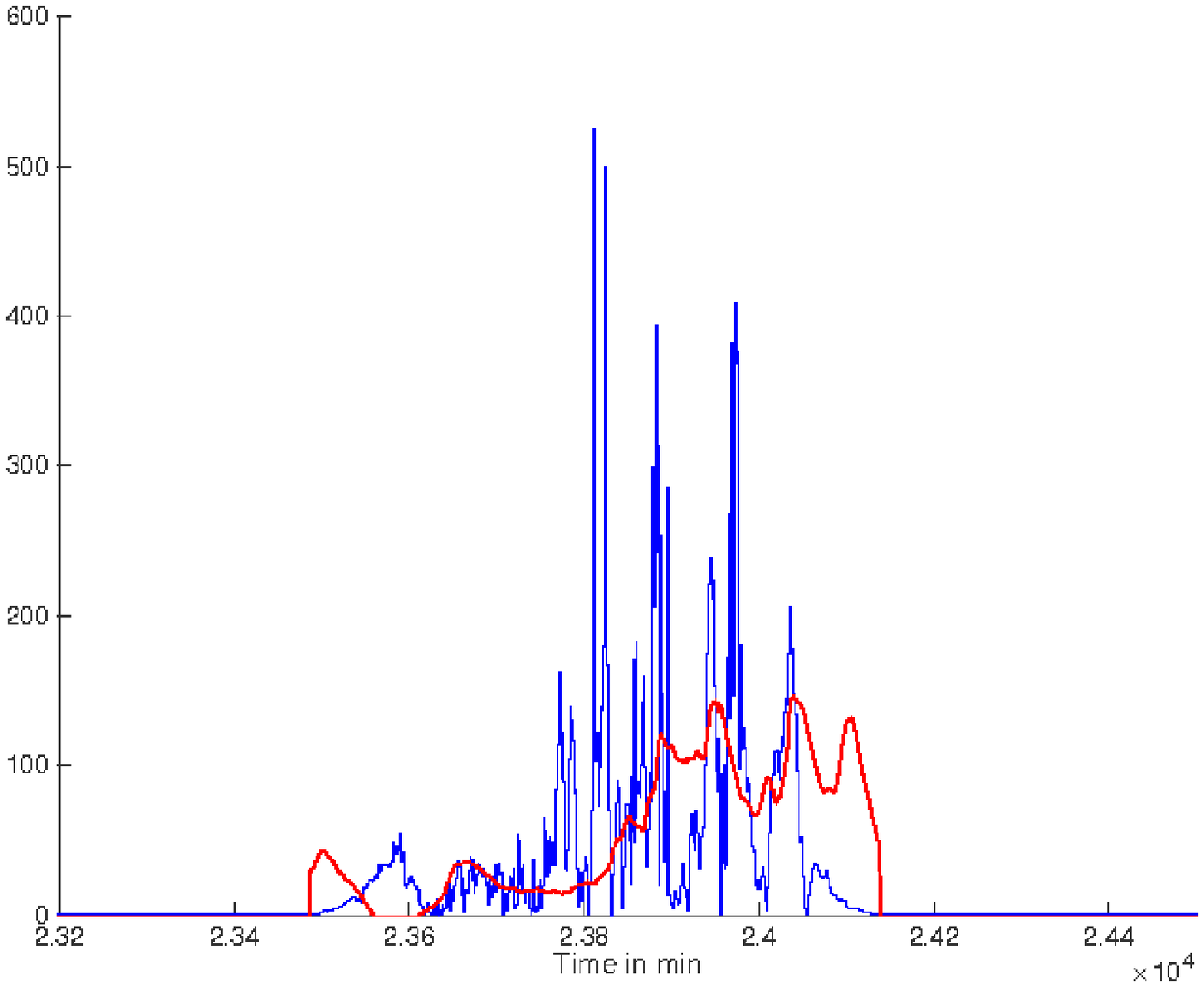}}}%
\subfigure[Zoom of \ref{figfv}-(a)]{
\resizebox*{6.05cm}{!}{\includegraphics{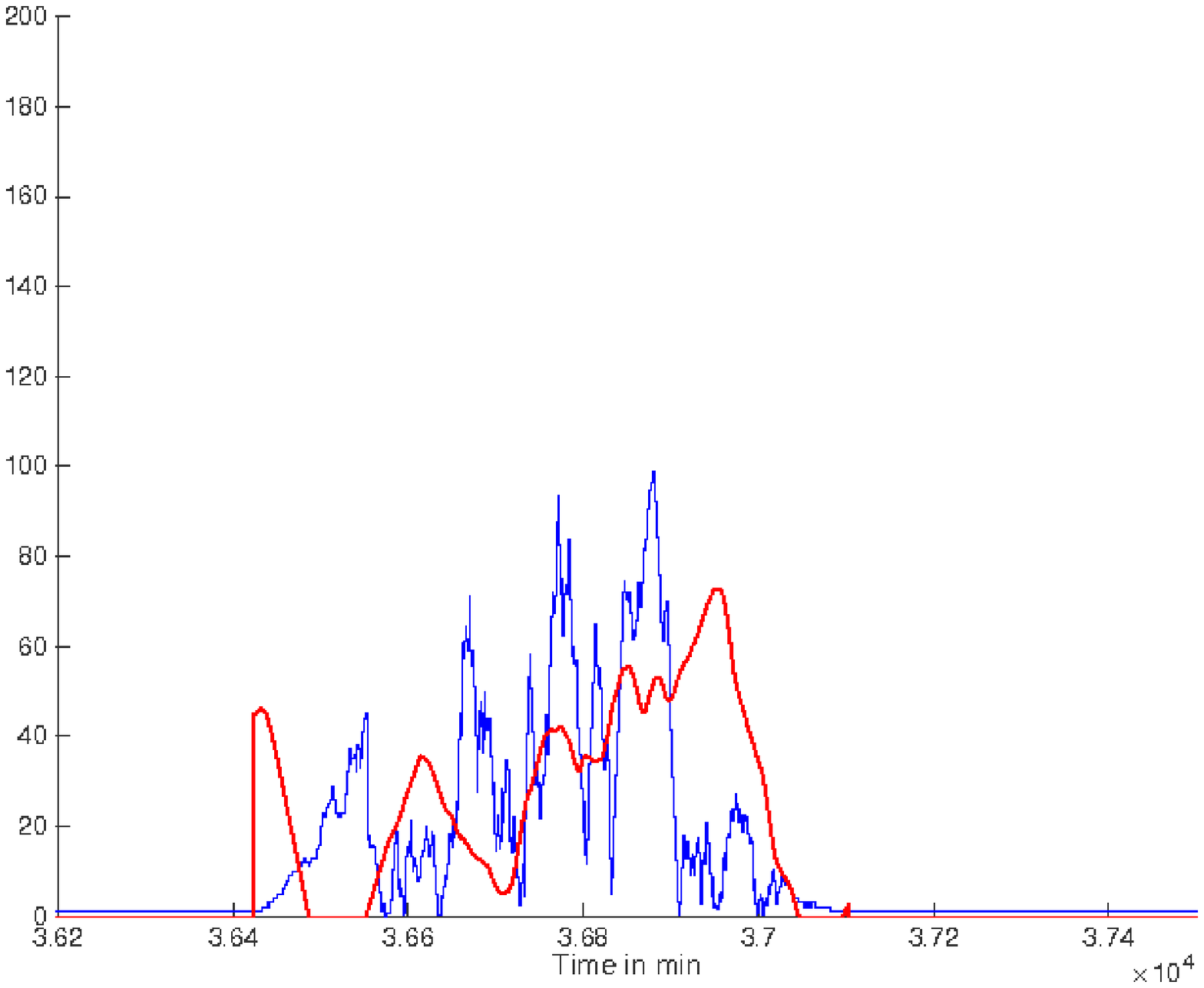}}}%
\caption{February : volatilty (blue) and its trend prediction (red)}%
\label{figfv}
\end{center}
\end{figure*}

\begin{figure*}
\begin{center}
\subfigure[Monthly view]{
\resizebox*{6.05cm}{!}{\includegraphics{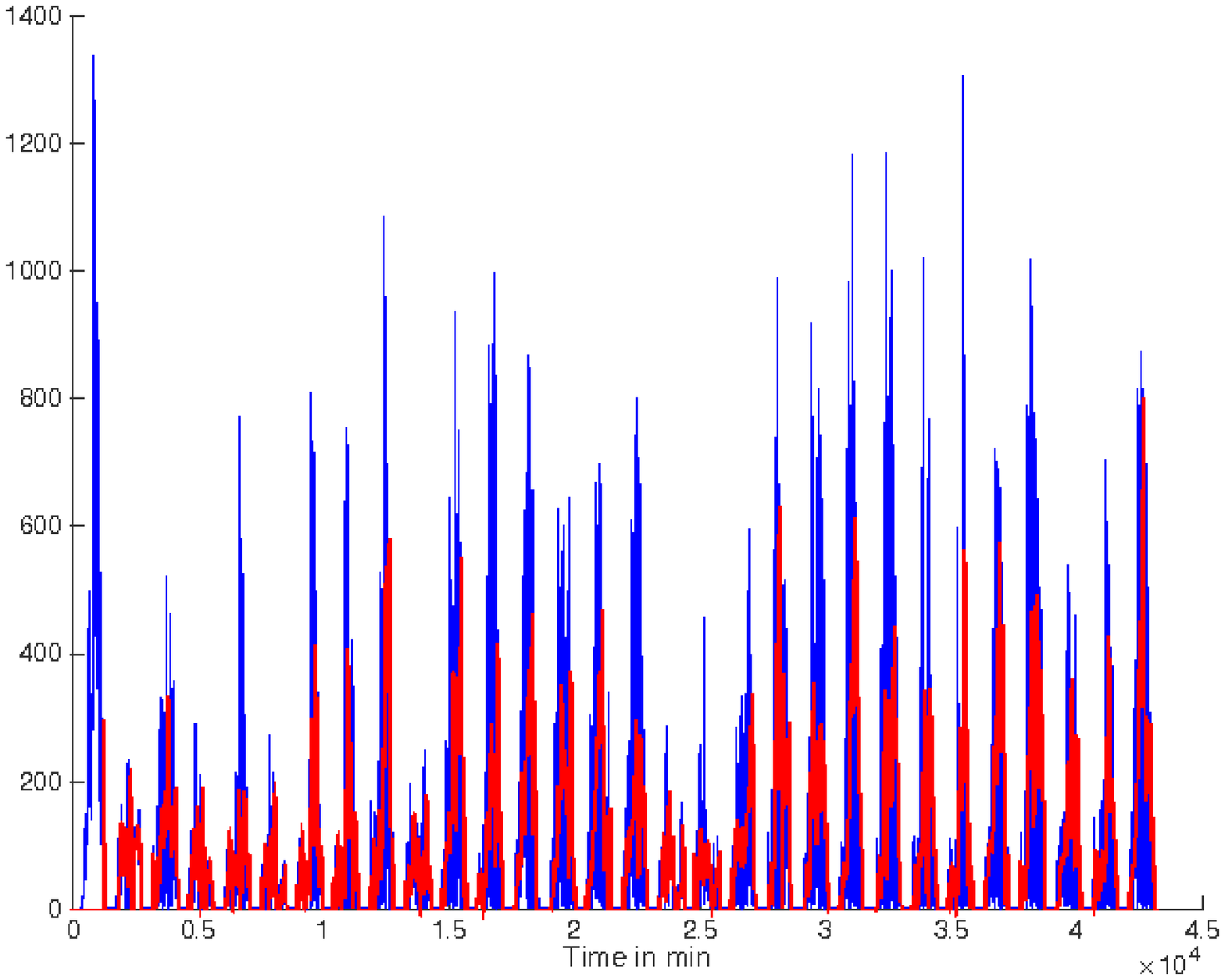}}}%
\subfigure[Zoom of \ref{figjv}-(a)]{
\resizebox*{6.05cm}{!}{\includegraphics{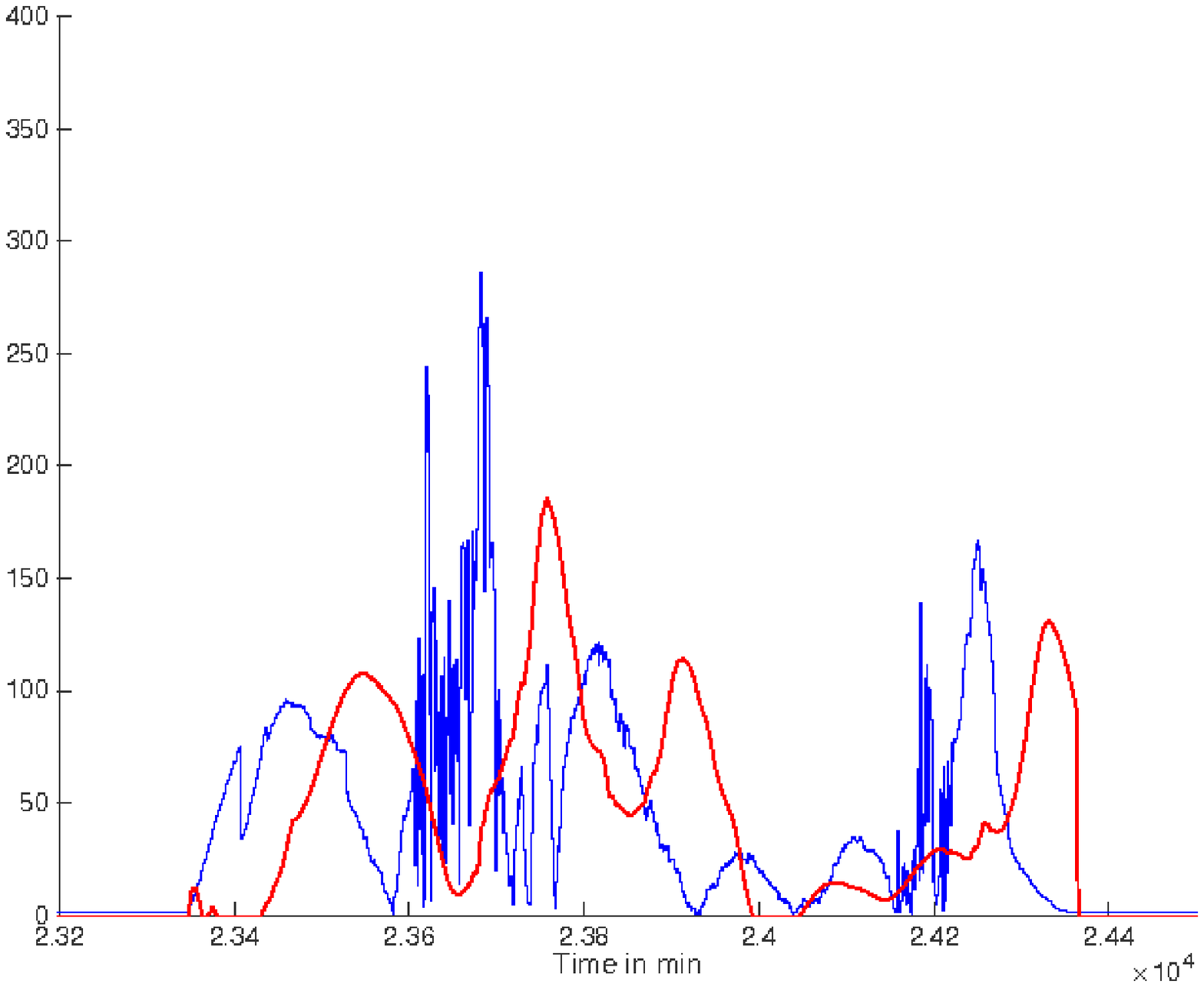}}}%
\subfigure[Zoom of \ref{figjv}-(a)]{
\resizebox*{6.05cm}{!}{\includegraphics{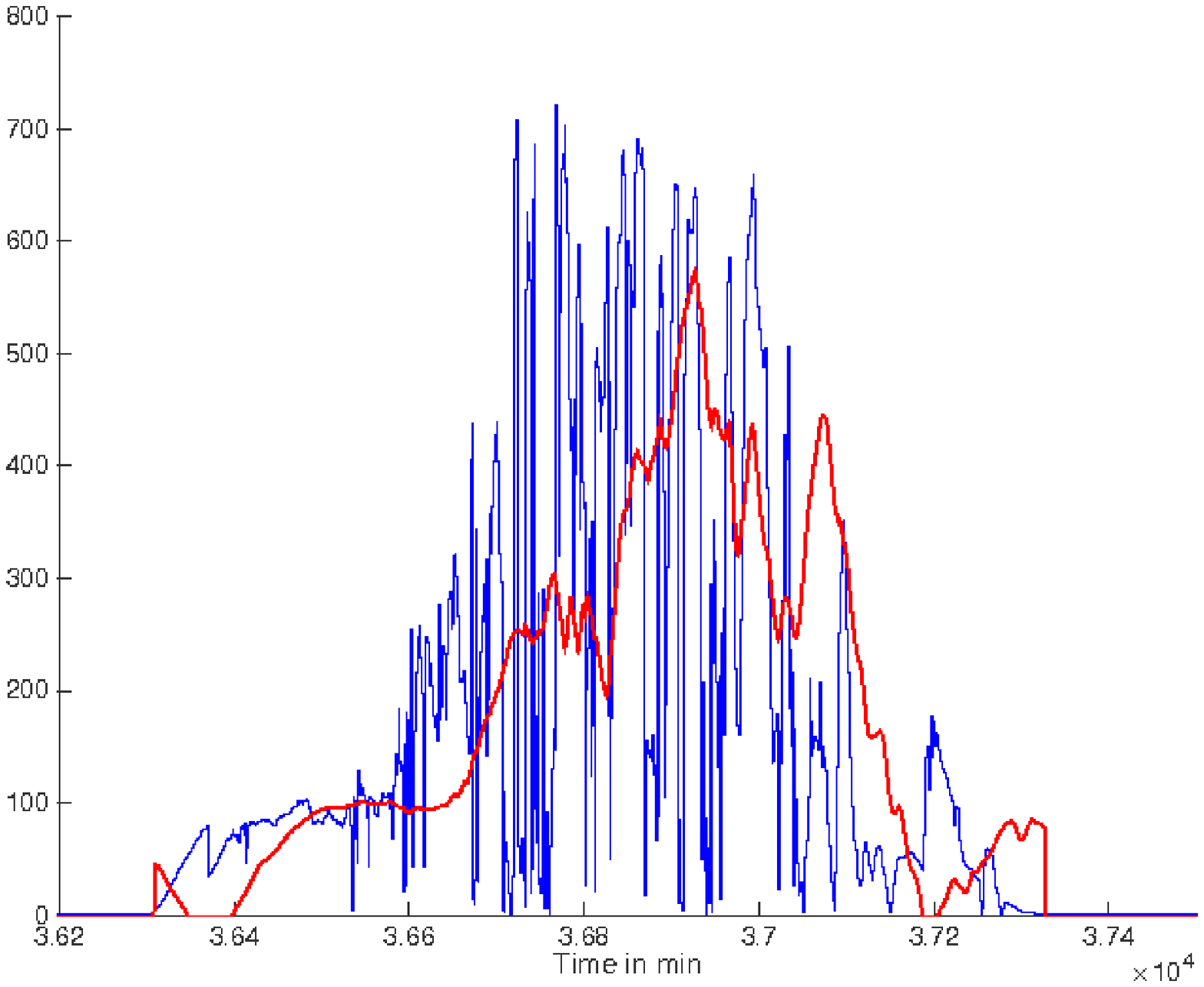}}}%
\caption{June : volatilty (blue) and its trend prediction (red)}%
\label{figjv}
\end{center}
\end{figure*}

Define the volatility via 
\begin{equation}\label{vol}
\text{{\bf Vol}}(t) = |X(t) - X_{\text{pred$60$}}(t - 60)|
\end{equation}
where $X_{\text{pred$60$}}(t - 60)$ is the forecast 1 h ahead of $X(t - 60)$. As stated before, it is obtained via the mean $E(X)(t)$ (see Equation \eqref{decomposition}). Figures \ref{figfv} and \ref{figjv} display the results. For simplicity's sake, set the following elementary persistence scheme for volatility
\begin{equation*}\label{volpred}
\text{{\bf Vol}}_{\text{pred$60$}}(t) = \text{{\bf Vol}}(t)
\end{equation*}
where $\text{{\bf Vol}}_{\text{pred$60$}}(t)$ is the forecast 1 h ahead of $\text{{\bf Vol}}(t)$.
\subsection{Normality tests}\label{normal}
\begin{figure*}
\begin{center}
\subfigure[February]{
\resizebox*{10.5cm}{!}{\includegraphics{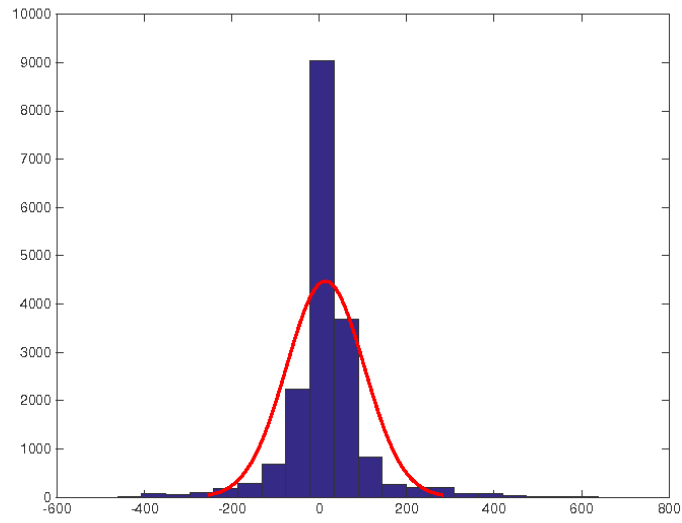}}}%
\\
\subfigure[June]{
\resizebox*{10.5cm}{!}{\includegraphics{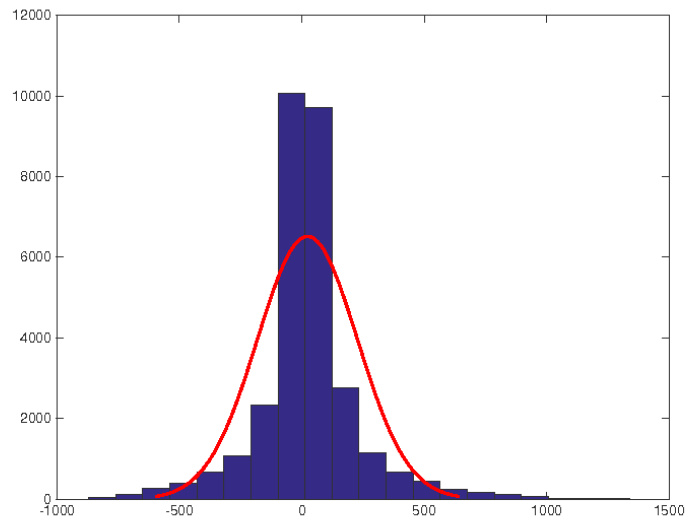}}}%
\caption{Signal distribution (blue) and the Gaussian distribution (red) }%
\label{H}
\end{center}
\end{figure*}

Let us associate to Equation \eqref{vol} the Equation 
$$
\text{Diff}(t) = X(t) - X_{\text{pred$60$}}(t - 60)
$$
Three classic normality tests (see, \textit{e.g.}, \cite{bourb,cryer,jarque,judge,thode}), namely
\begin{itemize}
\item Jarque-Bera,
\item Kolmogorov-Smirnov, 
\item Lilliefors,
\end{itemize}
reject the Gaussian property of the signal $\text{Diff}(t)$ for the twelve months of 2013. This is illustrated by Figures \ref{H}-(a) and \ref{H}-(b).

\subsection{Towards confidence bands}
\begin{figure*}
\begin{center}
\subfigure[Monthly view]{
\resizebox*{6.05cm}{!}{\includegraphics{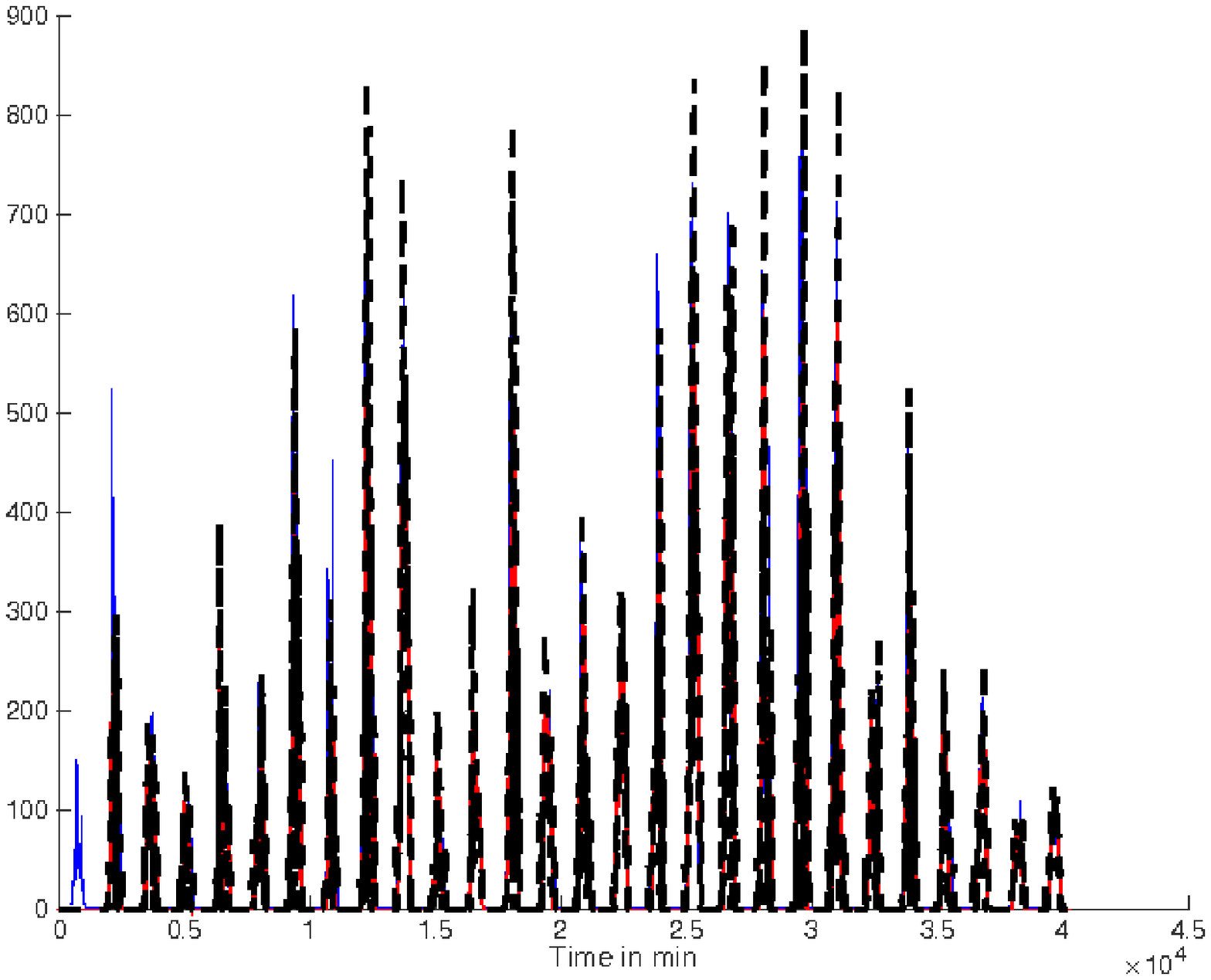}}}%
\subfigure[Zoom of \ref{figfi1}-(a)]{
\resizebox*{6.05cm}{!}{\includegraphics{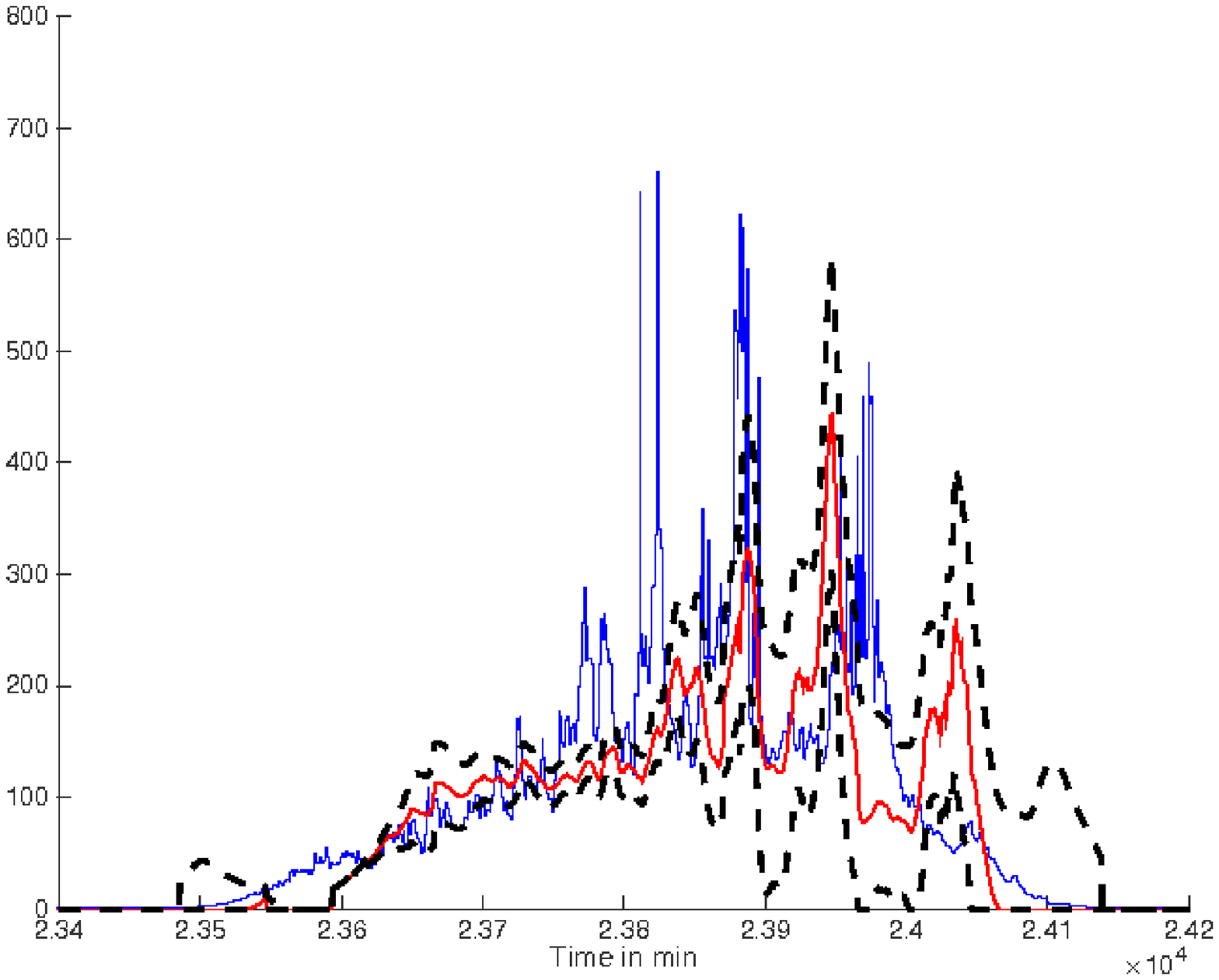}}}%
\subfigure[Zoom of \ref{figfi1}-(a)]{
\resizebox*{6.05cm}{!}{\includegraphics{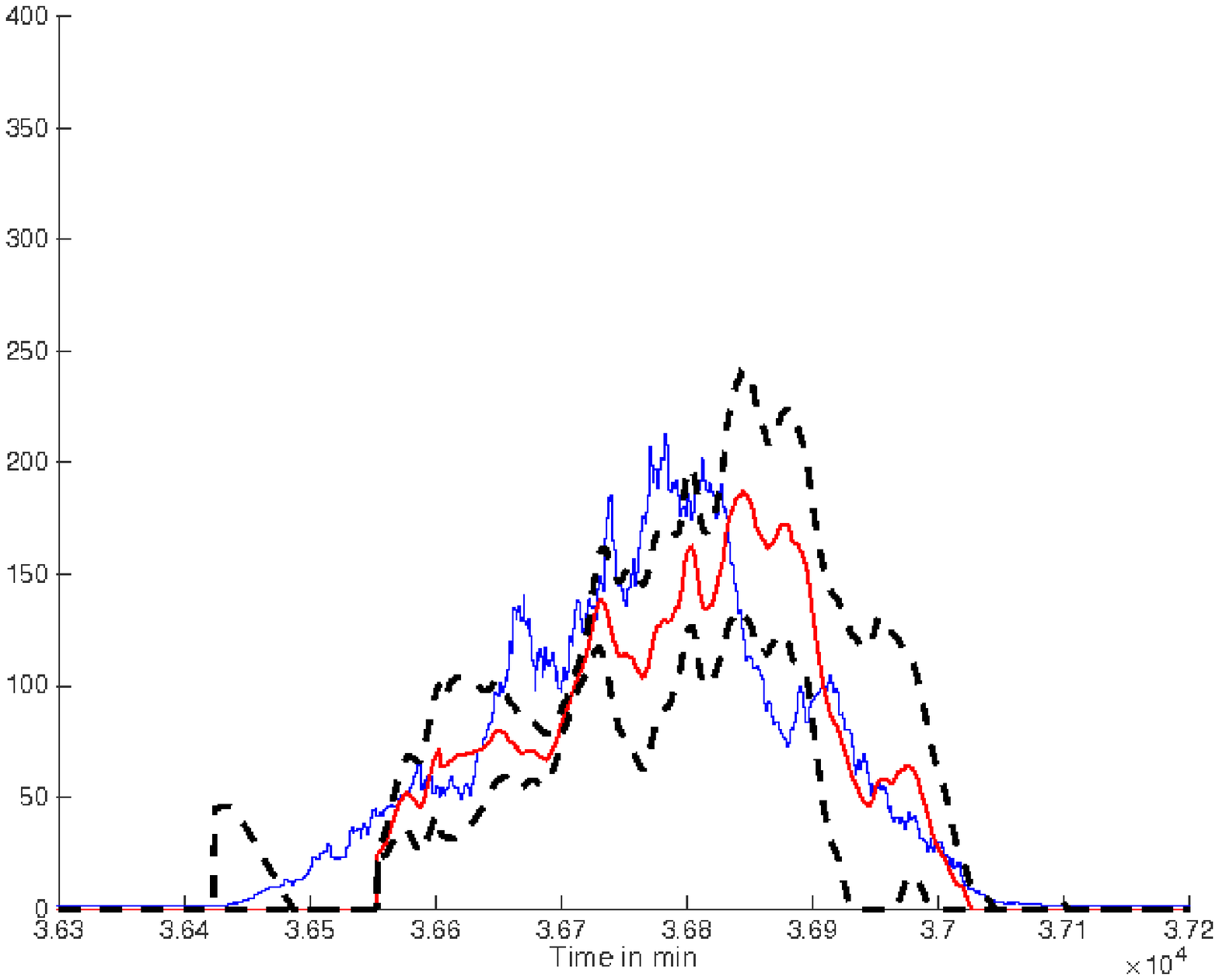}}}%
\caption{February: irradiance (blue), its prediction (red), confidence band
(black - -) (case: CB1)}%
\label{figfi1}
\end{center}
\end{figure*}

\begin{figure*}
\begin{center}
\subfigure[Monthly view]{
\resizebox*{6.05cm}{!}{\includegraphics{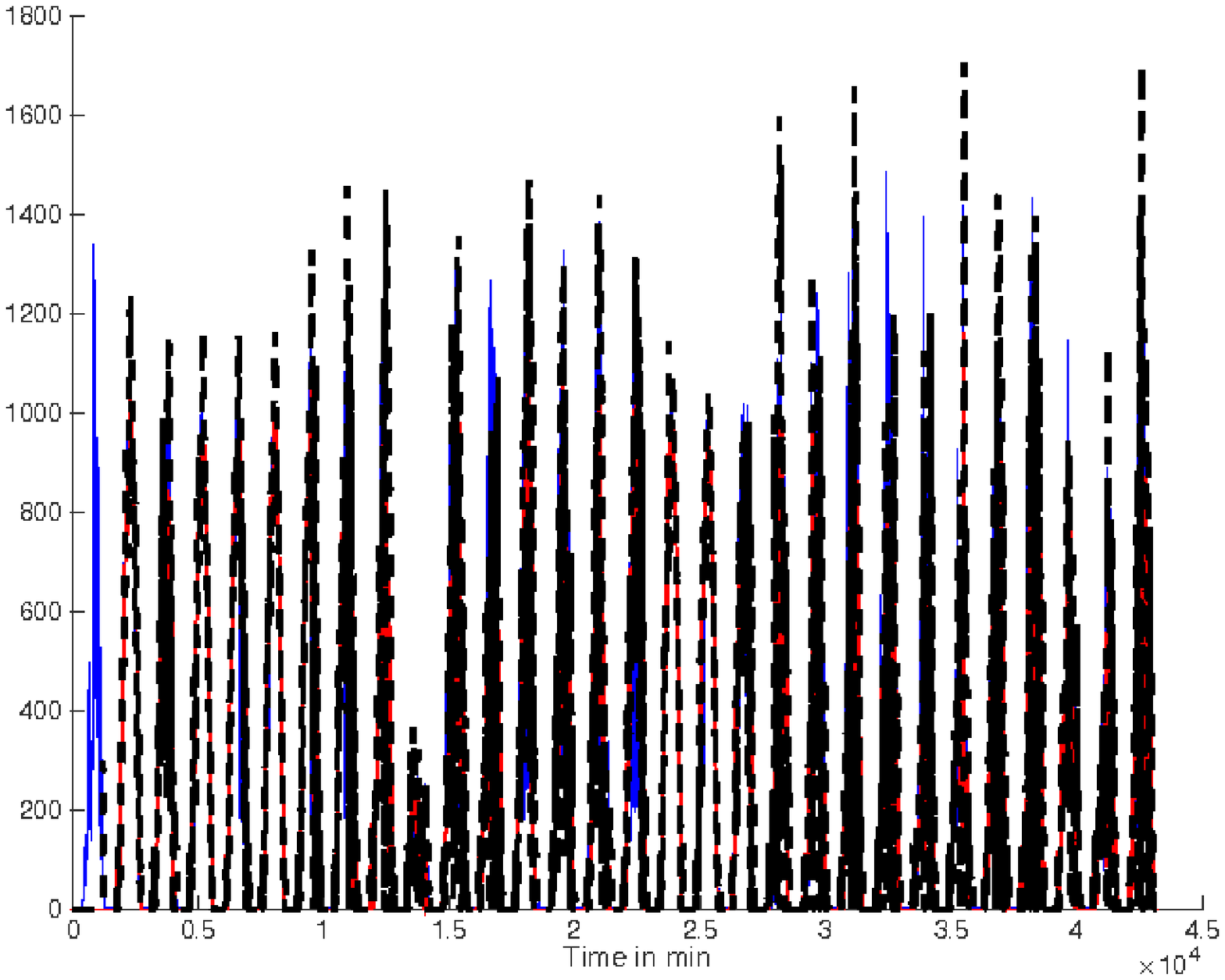}}}%
\subfigure[Zoom of \ref{figji1}-(a)]{
\resizebox*{6.05cm}{!}{\includegraphics{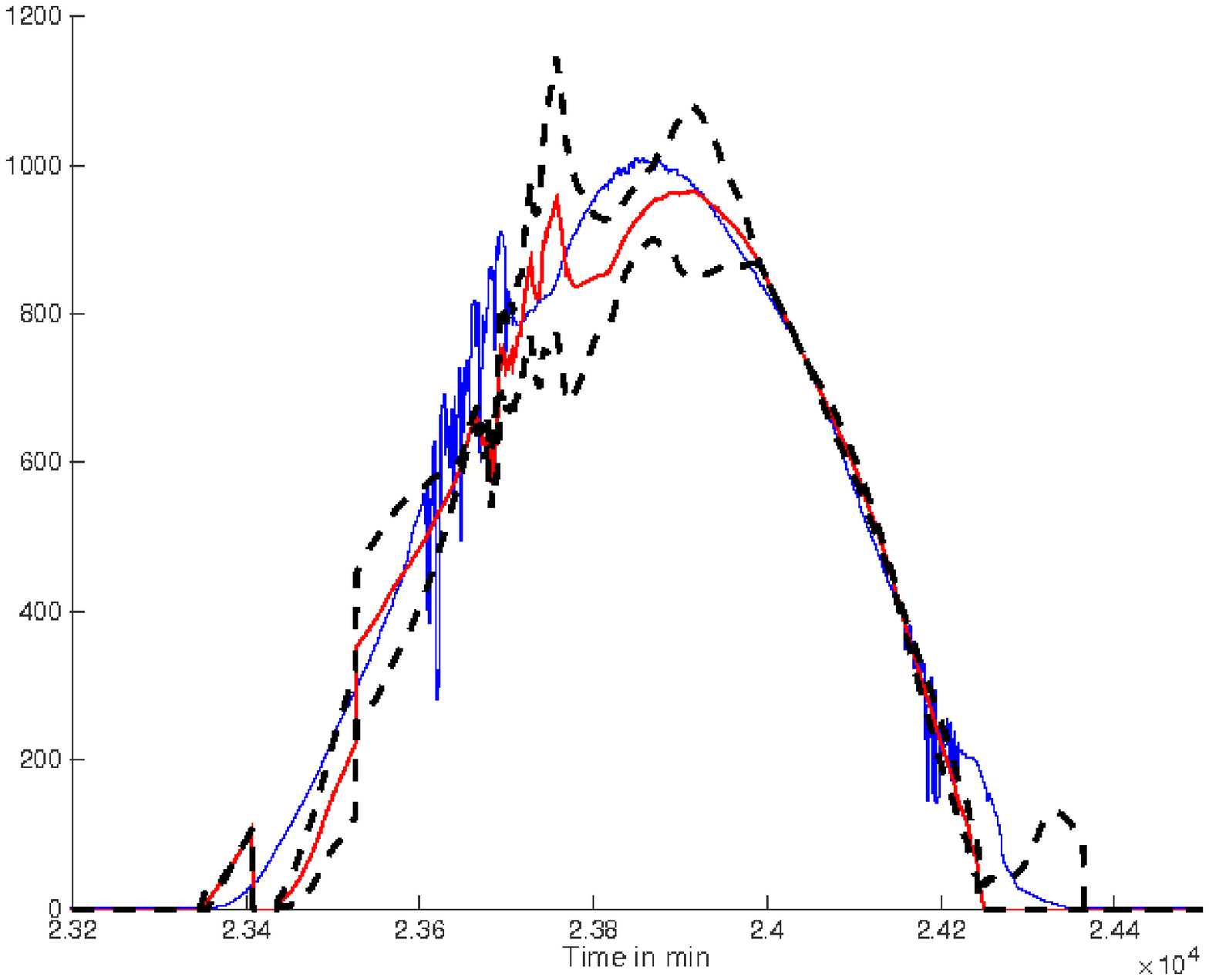}}}%
\subfigure[Zoom of \ref{figji1}-(a)]{
\resizebox*{6.05cm}{!}{\includegraphics{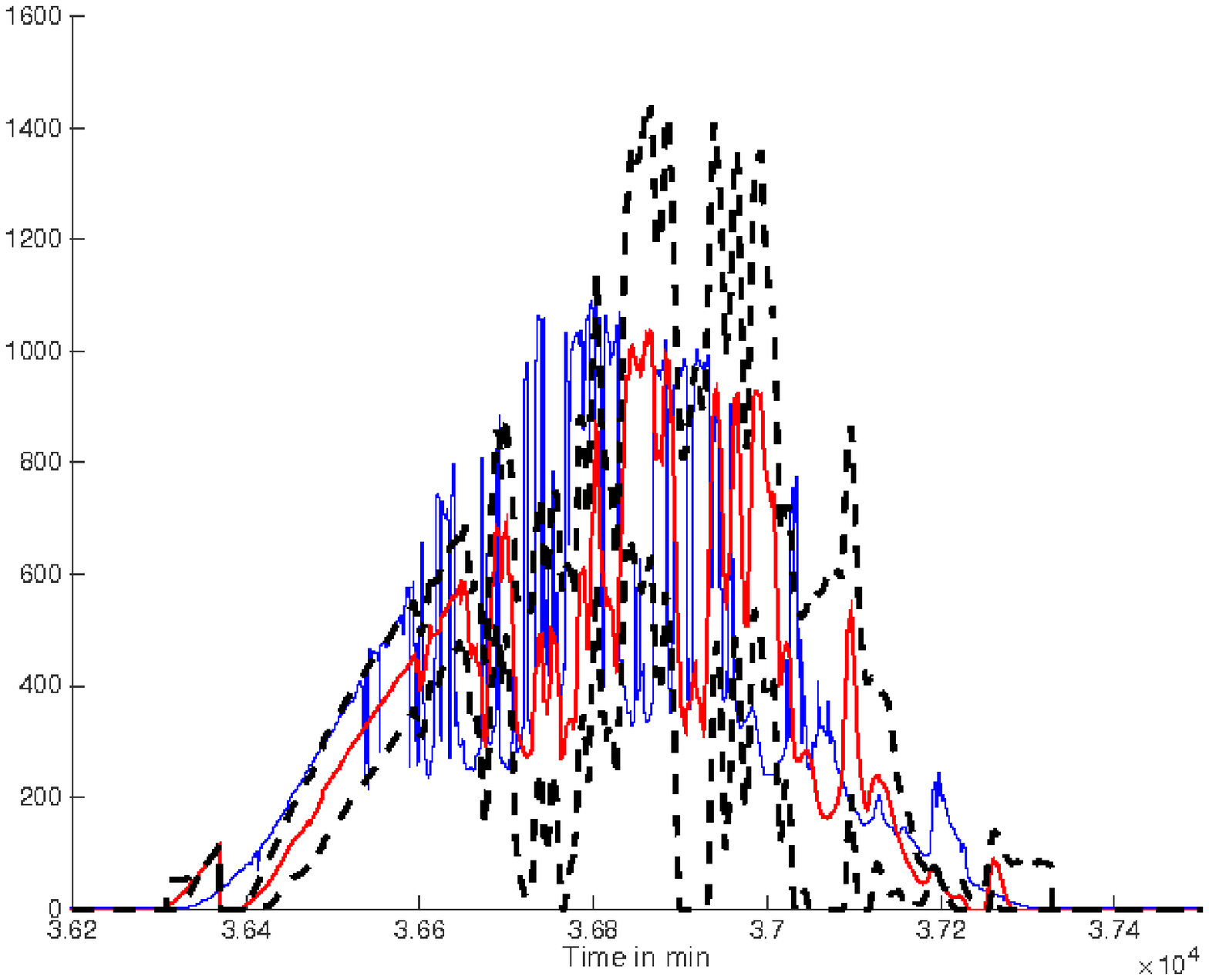}}}%
\caption{June: irradiance (blue), its prediction (red), confidence band (black
- -) (case: CB1)}%
\label{figji1}
\end{center}
\end{figure*}

Define a first \emph{confidence band} $\text{\bf CB}_{1, \  \text{pred$60$}}(t)$ by its frontiers
$$
X_{\text{pred$60$}}(t ) \pm \text{{\bf Vol}}_{\text{pred$60$}}(t)
$$
See Figures \ref{figfi1} and \ref{figji1}.

\begin{figure*}
\begin{center}
\subfigure[Monthly view]{
\resizebox*{6.05cm}{!}{\includegraphics{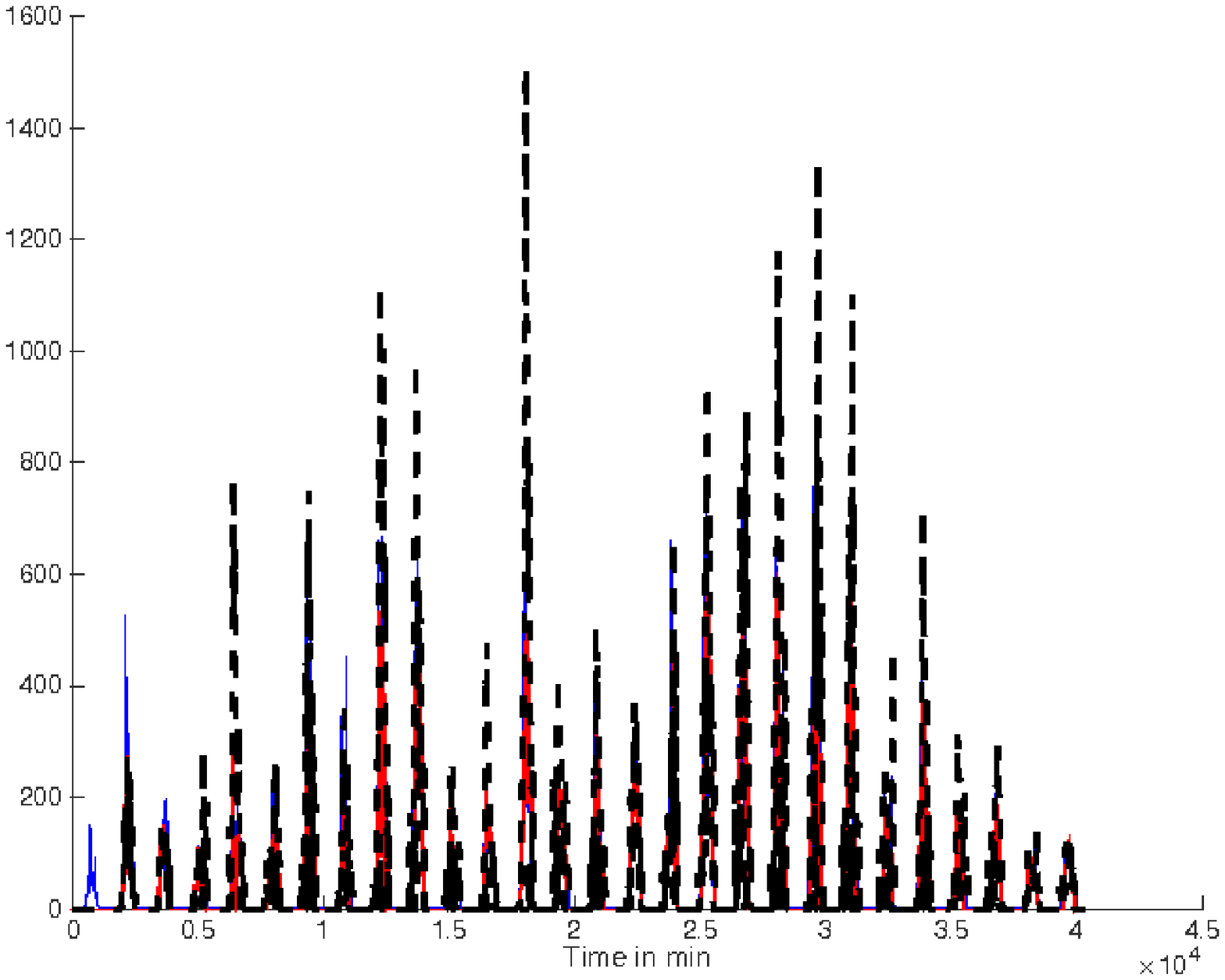}}}%
\subfigure[Zoom of \ref{figfi2}-(a)]{
\resizebox*{6.05cm}{!}{\includegraphics{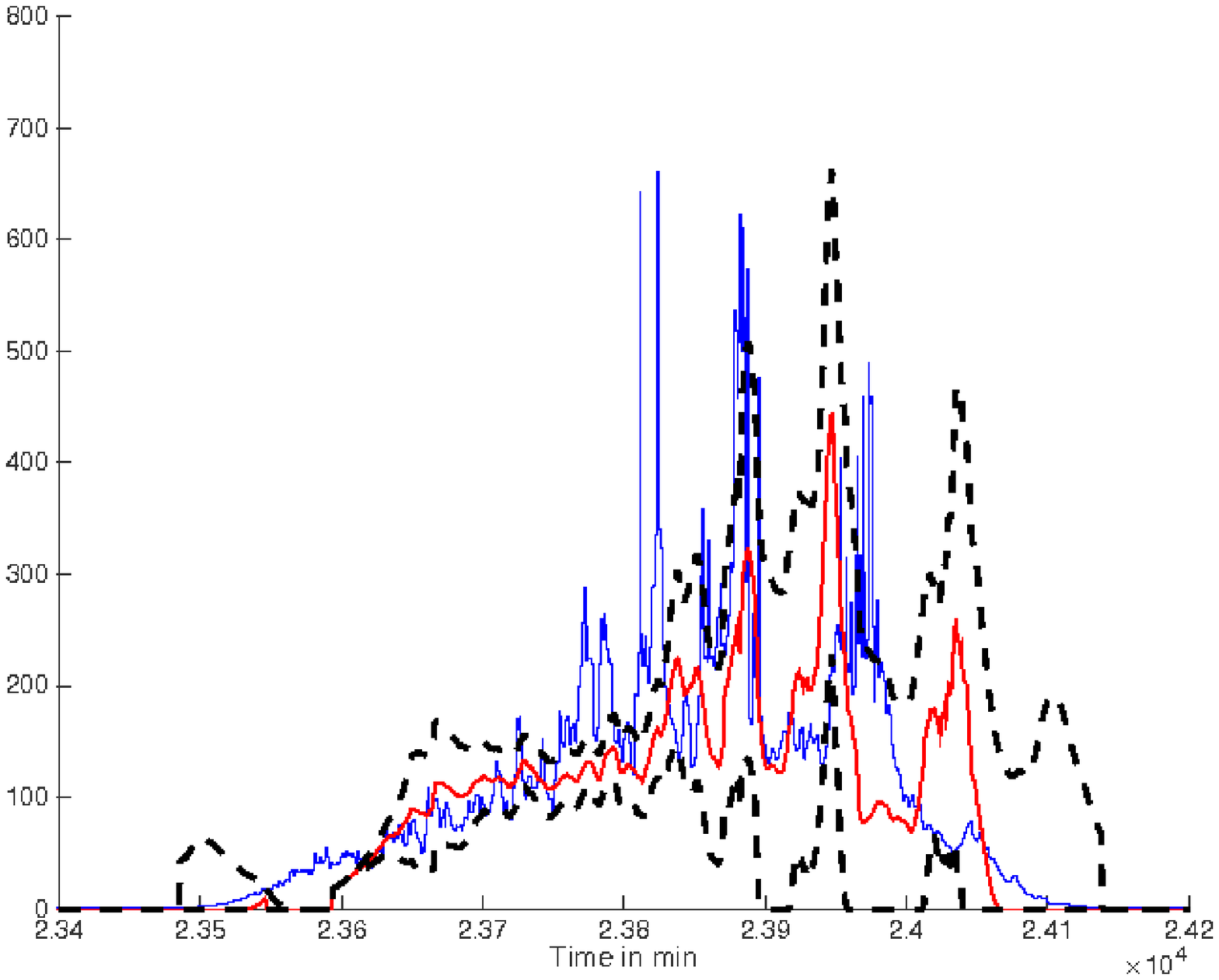}}}%
\subfigure[Zoom of \ref{figfi2}-(a)]{
\resizebox*{6.05cm}{!}{\includegraphics{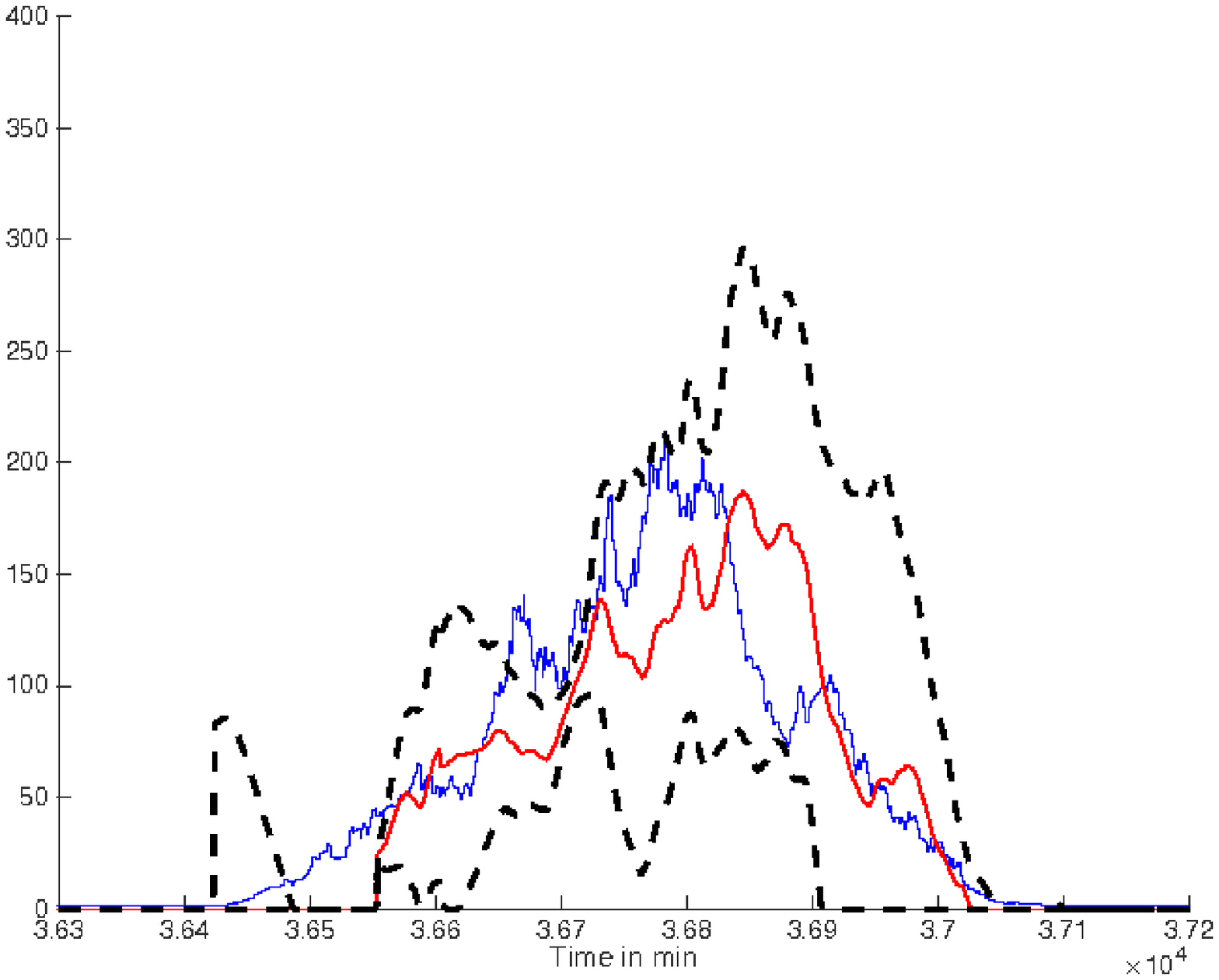}}}%
\caption{February: irradiance (blue), its prediction (red), confidence band
(black - -) (case: CB2)}%
\label{figfi2}
\end{center}
\end{figure*}

\begin{figure*}
\begin{center}
\subfigure[Monthly view]{
\resizebox*{6.05cm}{!}{\includegraphics{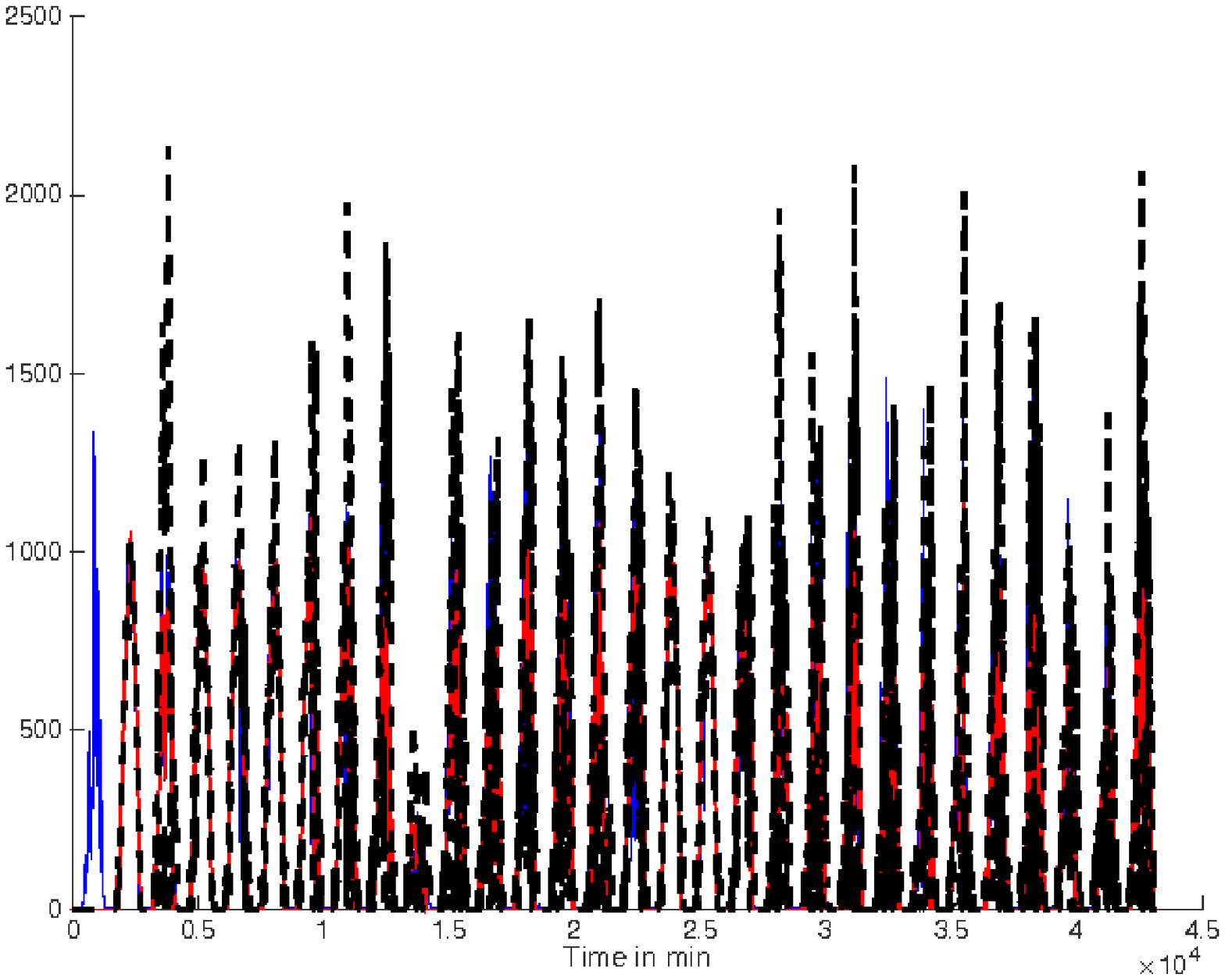}}}%
\subfigure[Zoom of \ref{figji2}-(a)]{
\resizebox*{6.05cm}{!}{\includegraphics{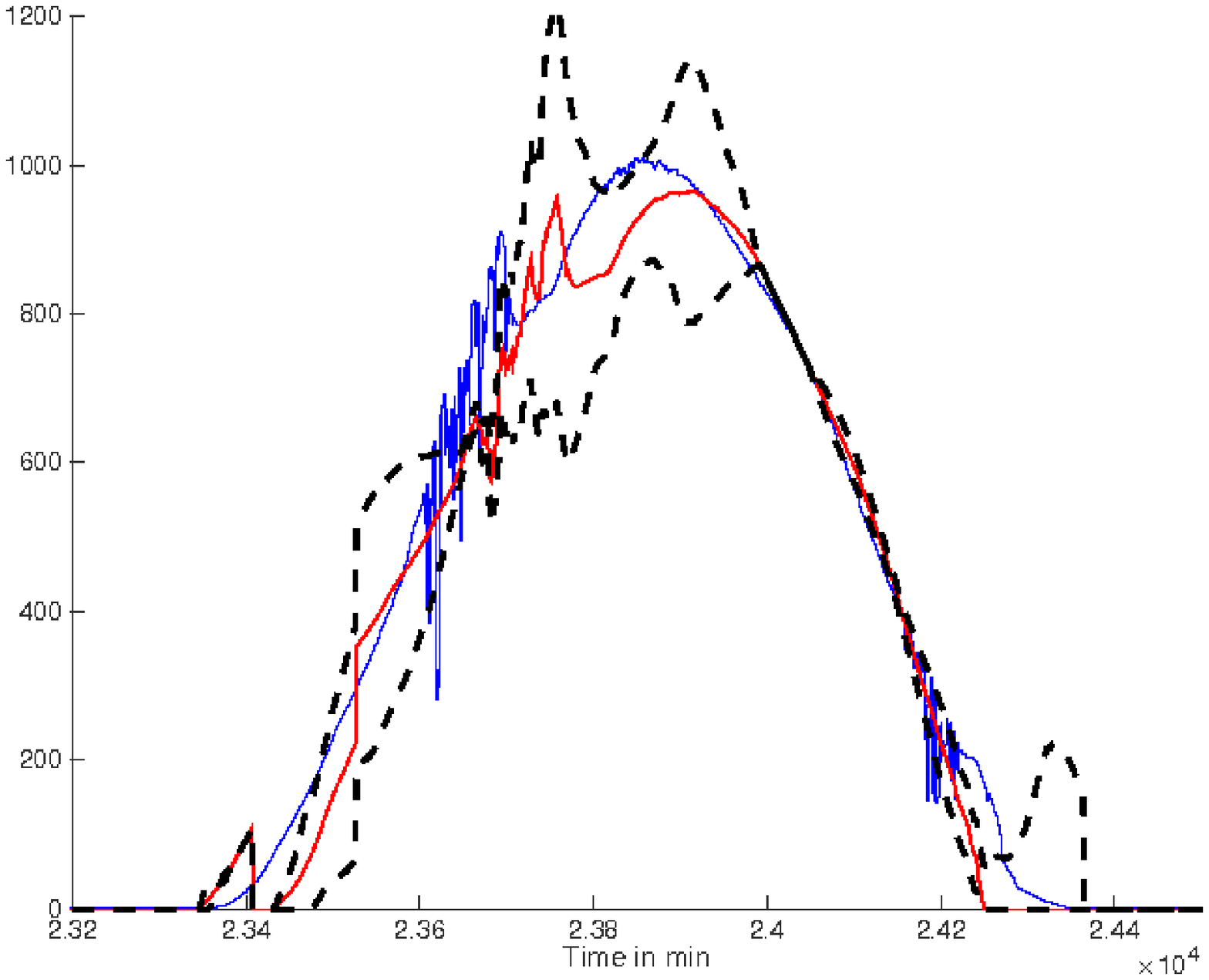}}}%
\subfigure[Zoom of \ref{figji2}-(a)]{
\resizebox*{6.05cm}{!}{\includegraphics{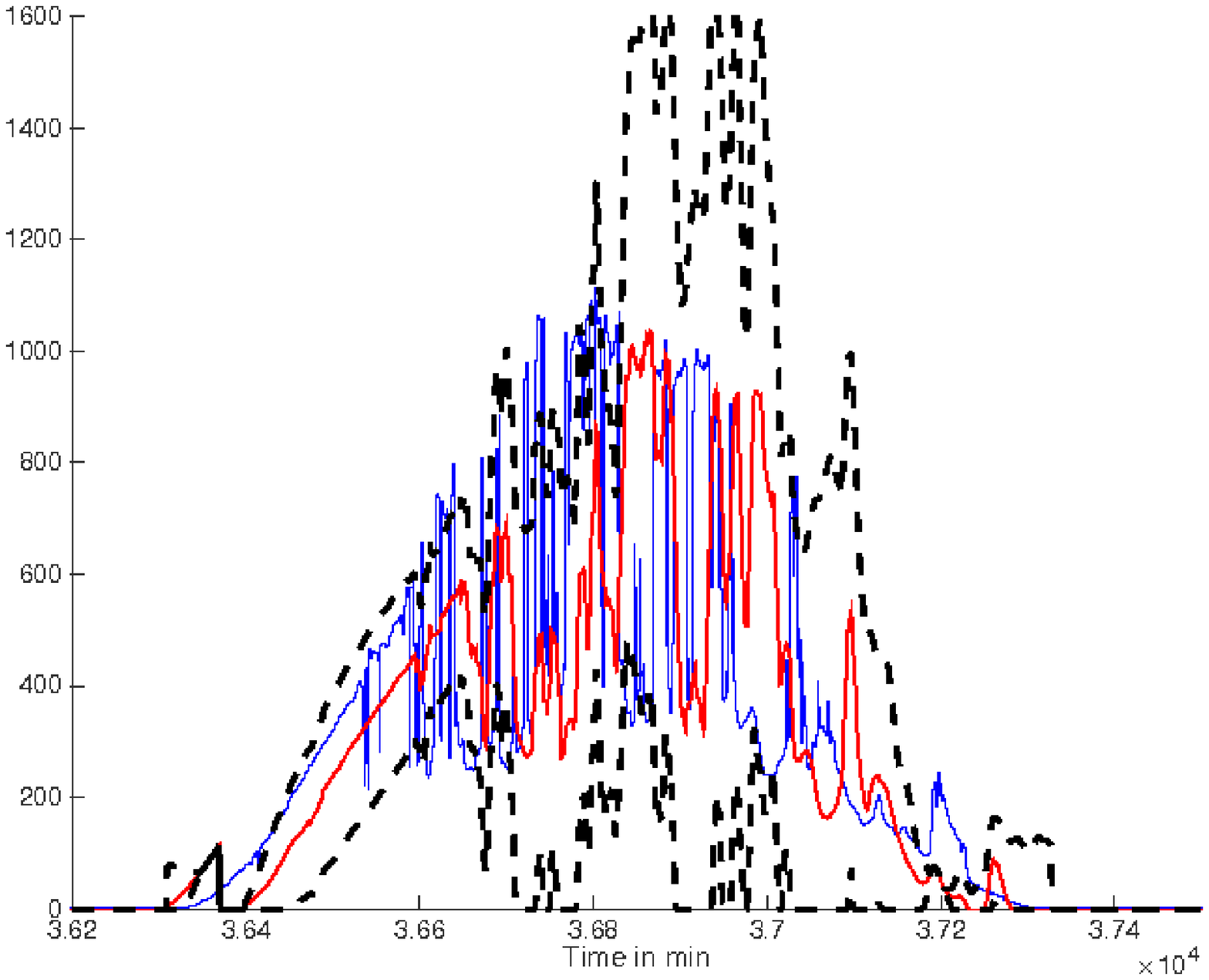}}}%
\caption{June: irradiance (blue), its prediction (red), confidence band (black
- -) (case: CB2)}%
\label{figji2}
\end{center}
\end{figure*}

In order to improve $\text{\bf CB}_{1, \  \text{pred$60$}}(t)$, define $\text{\bf CB}_{2, \  \text{pred$60$}}(t)$ by new frontiers 
$$
X_{\text{pred$60$}}(t ) \pm \alpha \text{{\bf Vol}}_{\text{pred$60$}}(t)
$$
where the parameter $\alpha$ is determined here by asking that during the three last days $68\%$ of the measured data were in $\text{\bf CB}_{2, \  \text{pred$60$}}(t)$.\footnote{The quantity $68\%$ is obviously inspired by 
the confidence intervals.}  See Figures \ref{figfi2} and \ref{figji2} .


\section{Conclusion}\label{conclusion}
Improving short-term forecasting and the corresponding confidence bands will be tackled in future publications. Seasonalities (see, \textit{e.g.}, \cite{ipag}) will of course play some r\^{o}le.

This communication shows that a ``good'' forecast (see, \textit{e.g.}, \cite{murphy}) should incorporate a measure of risk. Since, according to Section \ref{normal}, classic statistical confidence intervals are meaningless, we have introduced 
confidence bands, which do not necessitate any a priori probabilistic knowledge. Those bands will of course be further developed and applied to other domains. 

The fact that no probabilistic description is needed
has been already addressed in \cite{bruit,perp,agadir}. It is quite new in applied academic sciences, where a probabilistic description plays too often a key r\^{o}le. This fundamental epistemological issue
ought to be further developed (see also \cite{ayache}).

\begin{ack}
The installation of the meteorological station was made possible by the project  E2D2, or \textit{\'Energie, Environnement \& D\'eveloppement Durable}, which is supported by the \emph{Universit\'e de Lorraine} and
the \textit{R\'egion Lorraine}.
\end{ack}


\bibliography{ifacconf}             

\end{document}